\documentclass[12pt]{article}
\usepackage{amssymb,amscd,array}
\catcode `\@=11
\@addtoreset{equation}{section}

\newtheorem{thm}{Theorem}[section]

\def\qed{\blacksquare}
\newcommand{\be}{\begin{equation}}
\newcommand{\ee}{\end{equation}}
\newcommand{\bea}{\begin{eqnarray}}
\newcommand{\eea}{\end{eqnarray}}
\newcommand{\R}{\mathbb{R}}

\begin{document}
\begin{titlepage}

\begin{center}
{\bf \Large{Triangle (Causal) Distributions in the Causal Approach  \\}}
\end{center}
\vskip 1.0truecm
\centerline{D. R. Grigore, 
\footnote{e-mail: grigore@theory.nipne.ro}}
\vskip5mm
\centerline{Department of Theoretical Physics,}
\centerline{Institute for Physics and Nuclear Engineering ``Horia Hulubei"}
\centerline{Bucharest-M\u agurele, P. O. Box MG 6, ROM\^ANIA}

\vskip 2cm
\bigskip \nopagebreak
\begin{abstract}
\noindent
The tensor Feynman amplitudes are reduced to scalar integrals by a procedure of Passarino and Veltman. We provide an 
alternative approach based on the causal formalism.
\end{abstract}
\end{titlepage}

\section{Introduction}

One way to arrive at the Bogoliubov axioms of perturbative quantum field theory (pQFT) is by
analogy with non-relativistic quantum mechanics \cite{Gl}, \cite{H}; a discussion on this point can also be found in \cite{third order}.
We give the main ideas. Suppose that we have a time-dependent interaction potential
$V$. Then one goes to the interaction picture and the time evolution is governed by the evolution equation:
\be
{d \over dt}U(t,s) = - i V_{\rm int}(t) U(t,s); \qquad U(s,s) = I.
\ee

This equation can be solved in some cases by a perturbative method, namely the series
\be
U(t,s) \equiv \sum {(-i)^{n}\over n!} \int_{\R^{n}} dt_{1} \cdots dt_{n} T(t_{1},\dots,t_{n})
\ee
makes sense. The operators 
$
T_{n}(t_{1},\dots,t_{n})
$
are called  {\it chronological products}; $n$ is called the 
{\it order} of the perturbation theory. They verify a number of propertiesspelled in detail in the references from above.
Basically theay are unitarity and causality; the causality property means:
\bea
T_{n}(t_{1},\dots,t_{n}) = T_{m}(t_{1},\dots,t_{m})~T_{n-m}(t_{m+1},\dots,t_{n}),
\nonumber\\
{\rm for} 
\quad t_{j} > t_{k}, \quad j = 1,\dots,m; k = m+1,\dots,n.
\eea

An explicit formula is available (see the references above).

The purpose is to generalize this idea in the relativistic context especially the causality property. Essentially
we try to substitute
$
t \in \R
$
by a Minkowski variable 
$
x \in \R^{4}.
$
The chronological operators will be some operators
$
T(x_{1},\dots,x_{n})
$
and all the axioms from the non-relativistic case can be easily generalized rather naturally. 
The causally axiom is more subtle. We have to replace temporal succession
$
t_{1} > t_{2} 
$
by causal succession
$
x_{1} \succ x_{2}
$
which means that 
$
x_{1}
$
should not be in the past causal shadow of
$
x_{2}
$
i.e.
$
x_{2} \cap (x_{1} + \bar{V}^{+}) = \emptyset.
$
In formulas: if
$x_{i} \succ x_{j}, \quad \forall i \leq k, \quad j \geq k+1$
then we have:
\be
T(x_{1},\dots,x_{n}) =
T(x_{1},\dots,x_{k})~T(x_{k+1},\dots,x_{n}).
\label{causality1}
\ee
From here it follows that the ``initial condition"
$
T(x)
$
should satisfy
\be
[ T(x), T(y) ] = 0,\qquad (x - y)^{2} < 0
\ee
where for the Minkowski product we use the convention 
$
1,-1,-1,-1.
$
It a difficult problem to obtain solutions of the preceding equation. The solutions for pQFT are distribution-valued operators, 
(Wick monomials) and act in some Fock space where we can describe scattering processes with creation and annihilation of particles. 
Acording to Epstein and Glaser, we should solve directly the axioms of pQFT in an recursive way. 

So we start from Bogoliubov axioms \cite{BS}, \cite{EG} as presented in \cite{DF}, \cite{DB}; for every set of Wick polynomials 
$ 
A_{1}(x_{1}),\dots,A_{n}(x_{n}) 
$
acting in some Fock space
$
{\cal H}
$
one associates the operator-valued distributions
$ 
T^{A_{1},\dots,A_{n}}(x_{1},\dots,x_{n})
$  
called chronological products; it will be convenient to use another notation: 
$ 
T(A_{1}(x_{1}),\dots,A_{n}(x_{n})) 
$
and we should require skew-symmetry in all arguments: for arbitrary
$
A_{1}(x_{1}),\dots,A_{n}(x_{n})
$
we should have
\be
T(\dots,A_{i}(x_{i}),A_{i+1}(x_{i+1}),\dots,) =
(-1)^{f_{i} f_{i+1}} T(\dots,A_{i+1}(x_{i+1}),A_{i}(x_{i}),\dots)
\label{sqew}
\ee
where
$f_{i}$
is the number of Fermi fields appearing in the Wick monomial
$A_{i}$.

There are a number of rigorous ways to construct the chronological products:
(a) {\it Hepp axioms} \cite{H} (one rewrites the axioms in terms of vacuum averages of chronological products);
(b) {\it Polchinski flow equations} \cite{P}, \cite{S}  (one considers an ultra-violet cut-off
for the Feynman amplitudes and establishes some differential equations in this parameter);
(c) {\it The causal approach} due to Epstein and Glaser \cite{EG}, \cite{Gl}: is a recursive procedure for the basic objects
$ 
T(A_{1}(x_{1}),\dots,A_{n}(x_{n}))
$
and reduces the induction procedure to a distribution splitting of some distributions with causal support, or to the process of 
extension of distributions \cite{PS}. An equivalent point of view uses retarded products \cite{St1}. 
The causal method is the most elementary one from the point of view of conceptual clarity and also for practical computations.
It is a very good approach for the study of gauge models \cite{Sc1}, \cite{Sc2}.

The basic recursive idea of Epstein and Glaser starts from the chronological products
$$ 
T(A_{1}(x_{1}),\dots,A_{m}(x_{m})) \quad m = 1,2,\dots
$$
up to order 
$
n -1
$
and constructs a causal commutator in order $n$. 
For instance for
$
n = 2
$
the {\it causal commutator} according to:
\be
D(A(x),B(y)) = A(x)~B(y) - (-1)^{|A||B|}~B(y)~A(x)
\label{D2causal}
\ee
and after the operation of causal splitting one can obtain the second order 
chronological products. Generalizations of this formula are available for higher 
orders of the perturbation theory. In particular we have in the third order
\bea
D(A(x), B(y);C(z)) \equiv - [ \bar{T}(A(x), B(y)), C(z)]
\nonumber\\
+ (-1)^{|B||C|} [ T(A(x), C(z)), B(y)] + (-1)^{|A|(|B|+|C|)}  [ T(B(y), C(z)), A(x)]
\label{Dcausal}
\eea
where all commutators are understood to be graded. The causal commutators (\ref{D2causal}) and (\ref{Dcausal}) have the 
generic structure
\be
D = \sum d_{j}(X)~W_{j}(X)
\ee
where 
$
d_{j}(X)
$
are numerical distributions with causal support and 
$
W_{j}(X)
$
are Wick monomials. The numerical distributions
$
d_{j}
$
have various Lorentz indexes, so to compute them we need some sort of procedure which reduces everything to a certain master scalar
causal distribution. To obtain the corresponding chronological products one has to causally split only the master distribution.

A more popular approach is the so-called functional formalism; here one computes the chronological products making sense of
Feynman amplitudes. They are expressions of the type:
\bea
I_{N} \sim \int  {d^{4}l \over (2\pi)^{4}}~
{{\cal N}(l) \over \prod_{j = 1}^{N} [(l + q_{j-1})^{2} - m_{j}]}
\eea
which are associated to one-loop Feynman graphs \cite{EKMZ}. Here $N$ is the number of external particles and the denominator
$
{\cal N}(l) 
$
collects kinematic factors coming from vector and spinor propagators. Only the cases
$
N \leq 4
$
Then one is faced with the problem of computing integrals of the type
can produce ultra-violet divergences and a regularization is needed (usually the dimensional regularization.)

In the particular case of a triangle graph one needs 
to consider the regularized integrals of type $C$ (rel. (2.9) of \cite{EKMZ}). The idea is to use Lorentz covariance 
and express everything in terms of some scalar integrals. A recursive procedure due to Passarino and Veltman \cite{PV} is used.
In this procedure a singular region appears due to the annihilation of a certain Gram determinant. The procedure to circumvent
this singularity is to use different variables. For the general case more sophisticated methods are available \cite{EKMZ}.
The avoidance of the infra-red singularities is rather complicated in this approach. 

The purpose of this paper is to present how the computations are done in the framework of the causal approach. The idea 
is to compute some expressions with causal support properties called in \cite{EG} causal commutators. We will consider 
only the second and third order of perturbation theory. There causal commutators are sums of products between numerical
distributions with causal support and Wick monomials. The numerical distributions are similar to the type $C$ Feynman amplitudes 
from \cite{PV}, but no regularization procedure is needed. Also infra-red divergences do not appear because the chronological
products do not have such divergences: they appear only if we do the adiabatic limit. Finally, the treatment of the singularity
region associated to the Gram determinant seems to be easier.

We will present the computation of one-loop contributions in second and third order of perturbation theory 
in Sections \ref{second} and \ref{third}.

\newpage

\section{Second Order Distributions with Causal Support\label{second}}

In second order we have some typical distributions. 
We remind the fact that the Pauli-Villars distribution is defined by
\be
D_{m}(x) = D_{m}^{(+)}(x) + D_{m}^{(-)}(x)
\ee
where 
\be
D_{m}^{(\pm)}(x) = \pm {i \over (2\pi)^{3}}~
\int dp e^{i p\cdot x} \theta(\pm p_{0}) \delta(p^{2} - m^{2})
\ee
such that
\be
D^{(-)}(x) = - D^{(+)}(- x).
\ee

This distribution has causal support. In fact, it can be causally split
(uniquely) into an
advanced and a retarded part:
\be
D = D^{\rm adv} - D^{\rm ret}
\ee
and then we can define the Feynman propagator and anti-propagator
\be
D^{F} = D^{\rm ret} + D^{(+)}, \qquad \bar{D}^{F} = D^{(+)} - D^{\rm adv}.
\ee
All these distributions have singularity order
$
\omega(D) = -2
$.

These distributions do appear in the tree contributions to the chronological products. 

For  one-loop contributions in the second order we need the basic distributions
\be
d_{D_{1},D_{2}}(x) \equiv d^{(+)}_{D_{1},D_{2}}(x) + d^{(-)}_{D_{1},D_{2}}(x), \quad 
d^{(\pm)}_{D_{1},D_{2}}(x) \equiv \pm~{1 \over 2}~D_{1}^{(\pm)}(x)~D_{2}^{(\pm)}(x)
\label{d12}
\ee
(where
$
D_{j} \equiv D_{m_{j}}
$)
with causal support also. This expression is linear in
$
D_{1}
$
and
$
D_{2}
$.
We will also use the notation
\be
d_{12} \equiv d(D_{1},D_{2}) \equiv d_{D_{1},D_{2}}
\ee
and when no confusion about the distributions
$
D_{j} = D_{m_{j}}
$
can appear, we skip all indexes altogether. The causal split
\be
d_{12} = d_{12}^{adv} - d_{12}^{ret}
\ee
is not unique because
$
\omega(d_{12}) = 0
$
so we make the redefinitions
\be
d_{12}^{adv(ret)}(x) \rightarrow d_{12}^{adv(ret)}(x) + c~\delta(x)
\ee
without affecting the support properties and the order of singularity.
The corresponding Feynman propagators can be defined as above and will be
denoted as
$
d_{12}^{F}
$.

In \cite{loop} one can find the expressions of the dominant one-loop contributions from the chronological products. It is  
necessary to consider the case
$
D_{1} = D_{2} = D_{m}
$
and determine its Fourier transform. By direct computations it can be obtained
that
\be
\tilde{d}_{m,m}(k)
\equiv {1 \over (2\pi)^{2}} \int dx~ e^{i k\cdot x} d_{m,m}(x)
= - {1 \over 8 (2\pi)^{3}}~\varepsilon(k_{0})~\theta(k^{2} - 4 m^{2}) 
\sqrt{1 - {4 m^{2} \over k^{2}}}.
\label{d-mm}
\ee

We can consider associated causal distributions substituting in (\ref{d12}) 
$
D_{j} \rightarrow \partial_{\alpha}D_{j}
$
etc. It can be proved that we can reduce such causal distributed to some polynomials in partial derivatives applied to
$
d_{12}.
$
Detailed examples are provided in \cite{sr-gr}.

\newpage
\section{Third Order Causal Distributions of Triangle Type\label{third}}

First, we take
$
D_{j} = D_{m_{j}}, j = 1,2,3
$
and define
\bea
d_{D_{1},D_{2},D_{3}}(x,y,z) \equiv \bar{D}^{F}_{3}(x - y) 
[ D^{(-)}_{2}(z - x) D^{(+)}_{1}(y - z) - D^{(+)}_{2}(z - x) D^{(-)}_{1}(y - z)
]
\nonumber \\
+ D^{F}_{1}(y - z) 
[ D^{(-)}_{3}(x - y) D^{(+)}_{2}(z - x) - D^{(+)}_{3}(x - y) D^{(-)}_{2}(z - x)
]
\nonumber \\
+ D^{F}_{2}(z - x) 
[ D^{(-)}_{1}(y - z) D^{(+)}_{3}(x - y) - D^{(+)}_{1}(y - z) D^{(-)}_{3}(x - y)
]
\label{d123}
\eea
which are with causal support \cite{third order}. These distributions have the singularity order
$
\omega(d_{D_{1},D_{2},D_{3}}) = - 2
$.
As in the previous Section we use the alternative notation
\be
d_{123} \equiv  d(D_{1},D_{2},D_{3}) \equiv d_{D_{1},D_{2},D_{3}} 
\ee
and when there is no ambiguity about the distributions
$
D_{j}
$
we simply denote
$
d = d_{123}
$.
There are some associated distributions obtained from
$
d_{D_{1},D_{2},D_{3}}(x,y,z)
$
applying derivatives on the factors
$
D_{j} = D_{m_{j}}, j = 1,2,3
$
for instance
\bea
{\cal D}_{1}^{\mu}d_{D_{1},D_{2},D_{3}} \equiv
d_{\partial^{\mu}D_{1},D_{2},D_{3}},\quad
{\cal D}_{2}^{\mu}d_{D_{1},D_{2},D_{3}} \equiv
d_{D_{1},\partial^{\mu}D_{2},D_{3}},\quad
{\cal D}_{3}^{\mu}d_{D_{1},D_{2},D_{3}} \equiv
d_{D_{1},D_{2},\partial^{\mu}D_{3}},
\eea
and so on for more derivatives
$
\partial_{\alpha}
$
distributed on the factors
$
D_{j} = D_{m_{j}}, j = 1,2,3
$.

It is known that these distributions can be causally split in such a way that
the order of singularity, translation invariance and Lorentz covariance are
preserved. The same will be true for the corresponding Feynman distributions.
Because 
$
\omega(d_{123}) = - 2
$
and
$
\omega({\cal D}_{i}^{\mu}d_{123}) = - 1
$
the corresponding advanced, retarded and Feynman distributions are unique. For
more derivatives we have some freedom of redefinition.

As in the previous Section, let us consider the case
$
D_{1} = D_{2} = D_{3} = D_{m},~m > 0
$
and study the corresponding distribution
$
d_{m,m,m}.
$
We consider it as distribution in two variables
$
X \equiv x - z,\quad Y \equiv y - z
$
and we will need its Fourier transform which we define by
\be
\tilde{d}(p,q) \equiv  {1 \over (2\pi)^{4}}~\int e^{i (p\cdot X + q \cdot Y)}~d(X,Y). 
\ee
We will also need the distributions with causal support 
\bea
f_{1}(x,y,z) = \delta(y - z)~d_{m,m}(x - y)
\nonumber \\
f_{2}(x,y,z) = \delta(z - x)~d_{m,m}(y - z)
\nonumber \\
f_{3}(x,y,z) = \delta(x - y)~d_{m,m}(y - z)
\eea
with 
\be
\omega(f_{j}) = 0
\ee
and the Fourier transforms are:
\be
\tilde{f}_{1}(p,q) = {1 \over (2\pi)^{2}}~\tilde{d}_{m,m}(p),\quad
\tilde{f}_{2}(p,q) = {1 \over (2\pi)^{2}}~\tilde{d}_{m,m}(q),\quad
\tilde{f}_{3}(p,q) = {1 \over (2\pi)^{2}}~\tilde{d}_{m,m}(P)
\label{f}
\ee
with
$
P = p + q.
$

\newpage
\begin{thm}
The following formula is valid: 
\be
\tilde{d}_{m,m,m}(p,q) = {1\over 8 (2\pi)^{5}} {1 \over \sqrt{N}}~
[\epsilon(p_{0}) \theta(p^{2} - 4 m^{2})~ln_{1}
+ \epsilon(q_{0}) \theta(q^{2} - 4 m^{2})~ln_{2} 
+ \epsilon(P_{0}) \theta(P^{2} - 4 m^{2})~ln_{3} ]
\label{d-mmm}
\ee
where
\bea
ln_{1} \equiv ln\left({P\cdot q + \sqrt{N (1 - 4 m^{2}/p^{2})} \over
P\cdot q - \sqrt{N (1 - 4 m^{2}/p^{2})}}\right)
\nonumber \\
ln_{2} \equiv ln\left({P\cdot p + \sqrt{N (1 - 4 m^{2}/q^{2})} \over
P\cdot p - \sqrt{N (1 - 4 m^{2}/q^{2})}}\right)
\nonumber \\
ln_{3} \equiv ln\left({- p\cdot q + \sqrt{N (1 - 4 m^{2}/P^{2})} \over
- p\cdot q - \sqrt{N (1 - 4 m^{2}/P^{2})}}\right)
\eea
with the notations
$
P = p + q
$
and
$
N \equiv (p\cdot q)^{2} - p^{2} q^{2}.
$

The previous expression is continuous in the limit
$
N \rightarrow 0 ~(\Leftrightarrow p \parallel q)
$
and it is
\be
\tilde{d}_{m,m,m}(p,q) = 2(F_{1} + F_{2} + F_{3})
\ee
where
\be
F_{1} \equiv {1 \over P\cdot q}~\tilde{f}_{1}, \quad 
F_{2} \equiv {1 \over P\cdot p}~\tilde{f}_{2}, \quad
F_{3} \equiv {1 \over p\cdot q}~\tilde{f}_{3}. 
\ee
\end{thm}
{\bf Proof:} 
(i) From the definition (\ref{d123}) it follows that we have six contributions:
\be
d(X,Y) = \sum_{j=1}^{6}~d^{(j)}(X,Y)
\ee
of the form 
\be
d^{(j)}(X,Y) = d^{(j)}_{3}(X - Y)~d^{(j)}_{2}(- X)~d^{(j)}_{1}(Y),~j = 1,\dots,6
\ee
If we substitute
\be
d^{(j)}(X) =  {1 \over (2\pi)^{2}}~\int e^{- i k\cdot X}~\tilde{d}^{(j)}(k)
\ee
we get
\be
\tilde{d}^{(j)}(p,q) = {1 \over (2\pi)^{2}}~\int dk~\tilde{d}^{(j)}_{3}(k)~\tilde{d}^{(j)}_{2}(k - p)~\tilde{d}^{(j)}_{1}(k + q)
\ee
We consider for illustration the case 
$
j = 1
$
for which
\bea
\tilde{d}^{(1)}_{3}(k) = {1 \over (2\pi)^{2}}~{1 \over k^{2} - m^{2} - i~0}, 
\nonumber\\
\tilde{d}^{(1)}_{2}(k) = - {i \over 2\pi}~\theta( - k_{0})~\delta(k^{2} - m^{2}),\quad
\tilde{d}^{(1)}_{1}(k) = {i \over 2\pi}~\theta(k_{0})~\delta(k^{2} - m^{2}).
\eea
We substitute in the previous formula and obtain
\be
\tilde{d}^{(1)}(p,q) = {1 \over (2\pi)^{6}}~\int dk {1 \over k^{2} - m^{2} - i~0}~
\theta(p_{0} - k_{0})~\delta((p - k)^{2} - m^{2})~\theta(k_{0} + q_{0})~\delta((k + q)^{2} - m^{2})
\label{d1}
\ee
We make the change of variables 
$
k \rightarrow k + p
$
leading to
\be
\tilde{d}^{(1)}(p,q) = {1 \over (2\pi)^{6}}~\int dk {1 \over (k + p)^{2} - m^{2} - i~0}~
\theta(- k_{0})~\delta(k^{2} - m^{2})~\theta(k_{0} + P_{0})~\delta((k + P)^{2} - m^{2})
\label{d1a}
\ee
and afterwards we use the distribution 
$
\delta(k^{2} - m^{2})
$
to integrate over
$
k_{0}.
$
The result is 
\be
\tilde{d}^{(1)}(p,q) = {1 \over (2\pi)^{6}}~\int_{\omega_{\bf k} \leq P_{0}} {d{\bf k} \over 2 \omega_{\bf k}}~
\delta(P^{2} - 2 P_{0}\omega_{\bf k} - 2 {\bf P}\cdot {\bf k})~
(p^{2} - 2 p_{0}\omega_{\bf k} - 2 {\bf p}\cdot {\bf k} - i~0)^{-1}
\label{d1b}
\ee
where we have defined
$
\omega_{\bf k} \equiv \sqrt{{\bf k}^{2} + m^{2}}.
$

This expression is Lorentz invariant. We can use this fact to prove that the integral is zero in the cases 
$
P^{2} \leq 0
$
and
$
P^{2} > 0, P_{0} < 0.
$
We are left with the case 
$
P^{2} = M^{2}~ (M > 0), P_{0} \geq 0
$
so we can evaluate it in a frame where
$
P = (M,{\bf 0}).
$
In this frame we get
\be
\tilde{d}^{(1)}(p,q) = {1 \over (2\pi)^{6}}~\int_{\omega_{\bf k} \leq M} {d{\bf k} \over 2 M^{2}}~
\delta\Bigl(\omega_{\bf k} - {M\over 2}\Bigl)~
(p^{2} - M p_{0} - 2 {\bf p}\cdot {\bf k} - i~0)^{-1}
\label{d1c}
\ee

It is obvious that we must consider two cases:
$
{\bf p} \not= {\bf 0}
$
and
$
{\bf p} = {\bf 0}.
$

(ii) We first consider the case 
$
{\bf p} \not= {\bf 0}.
$
We perform the integration in spherical coordinates 
$
(r, \theta,\phi)
$
with the third axis
$
{\bf e}_{3} \parallel {\bf p}.
$
The integrals over
$
\phi
$
and $r$ are elementary. In particular we find out that the integral is non-zero only if
$
M \geq 2m
$
and we are left with
\be
\tilde{d}^{(1)}(p,q) = \theta(P_{0})~\theta(P^{2} - 4m^{2})~{r_{0} \over 4(2\pi)^{5}M}~
\int d\theta sin\theta~(p^{2} - M p_{0} - 2 |{\bf p}|r_{0} cos\theta - i~0)^{-1}
\label{d1d}
\ee
where
$
r_{0} \equiv \sqrt{{M^{2}\over 4} - m^{2}}
$. 
With the new variable
$
z = cos\theta
$
we get
\be
\tilde{d}^{(1)}(p,q) = \theta(P_{0})~\theta(P^{2} - 4m^{2})~{r_{0} \over 4(2\pi)^{5}M}~I_{0}(A,B)
\label{d1e}
\ee
where
\be
I_{0}(A,B) \equiv \int_{-1}^{1} {dz \over A - B z}
\ee
and
\be
A = p^{2} - M p_{0} - i~0,\quad B = 2 |{\bf p}|r_{0}.
\ee

The integral is elementary
\be
I_{0}(A,B) = {1 \over B}~ln\Bigl({A + B\over A - B}\Bigl).
\ee
Now we want to rewrite the expression 
$
\tilde{d}^{(1)}(p,q)
$
in covariant coordinates. We will use the invariant $N$ defined in the statement of the theorem and also
$
I = P\cdot p.
$
In the particular frame we have used we have
$
I = M~p_{0}, \quad N = M^{2} {\bf p}^{2}
$
so it follows that we also have in this frame
$
A = - p\cdot q, r_{0} = \sqrt{{P^{2}\over 4} - m^{2}}, {r_{0} \over B} = \sqrt{P^{2}\over N}. 
$
So, the formula
\be
\tilde{d}^{(1)}(p,q) = \theta(P_{0})~\theta(P^{2} - 4m^{2})~{1 \over 8(2\pi)^{5}}~{1 \over \sqrt{N}} ln_{3} 
\label{d1f}
\ee
is valid in the particular frame and, because of Lorentz invariance, it is valid in general.

Next we use the relation
\be
\tilde{d}^{(2)}(p,q) = -\tilde{d}^{(1)}(- q,- p) 
\label{d2}
\ee
and the obtain the other piece proportional to 
$
ln_{3}.
$

In a similar way we obtain
\be
\tilde{d}^{(3)}(p,q) = - \tilde{d}^{(1)}(q,- P)^{*} 
\label{d3}
\ee
\be
\tilde{d}^{(4)}(p,q) = - \tilde{d}^{(3)}(- p,- q) 
\label{d4}
\ee
and these relations lead to the 
$
ln_{1}
$
contribution. Finally
\be
\tilde{d}^{(5)}(p,q) = \tilde{d}^{(3)}(q,p) 
\label{d5}
\ee
\be
\tilde{d}^{(6)}(p,q) = \tilde{d}^{(4)}(q,p) 
\label{d6}
\ee
and these relations lead to to the 
$
ln_{2}
$
contribution. 

(iii) We consider now the case 
$
{\bf p} = {\bf 0}.
$
We return to (\ref{d1c}) which is in this case 
\be
\tilde{d}^{(1)}(p,q) = {1 \over (2\pi)^{6}}~\int_{\omega_{\bf k} \leq M} {d{\bf k} \over 2 M^{2}}~
\delta\Bigl(\omega_{\bf k} - {M\over 2}\Bigl)~(p^{2} - M p_{0} - i~0)^{-1}
\label{d1g}
\ee
We also perform the integration in spherical coordinates, but now we can chose the axis
$
{\bf e}_{3}
$
at will. The result is similar to (\ref{d1e}):
\be
\tilde{d}^{(1)}(p,q) = \theta(P_{0})~\theta(P^{2} - 4m^{2})~{r_{0} \over 2(2\pi)^{5}M A}.
\label{d1h}
\ee

(iv) We prove now that the expression (\ref{d1e}) is continuous in the limit 
$
{\bf p} \rightarrow {\bf 0}
$
and gives us the preceding formula. This is in fact, equivalent to 
\be
lim_{B \rightarrow 0} I_{0}(A,B) = {2 \over A}
\ee
and this is elementary. Lastly, we give the covariant form of (\ref{d1h}). As in the previous case we have:
\be
\tilde{d}^{(1)}(p,q) = {2 \over (2\pi)^{2}}~{1 \over p\cdot q}~\tilde{d}_{m,m}^{(+)}(P) 
\label{d1h)}
\ee
where the expression
$
\tilde{d}_{m,m}
$
was defined in the previous section. We obtain the formula from the statement.
$\qed$
\newpage

We proceed in the same way for the distributions
\be
d_{i}^{\mu} \equiv {\cal D}_{i}^{\mu} d
\ee
and we have
\be
\omega(d_{j}^{\mu}) = - 1
\label{deg1}
\ee
and the result is
\begin{thm}
For 
$
N \not= 0
$
the following formula is true:
\be
\tilde{d}_{3}^{\mu}(p,q) = i~({\cal A}^{\mu}_{1}~\tilde{d} + {\cal A}^{\mu}_{2}~\tilde{f}_{3} + 
{\cal A}^{\mu}_{3}~\tilde{f}_{1} + {\cal A}^{\mu}_{4}~\tilde{f}_{2} )
\label{d3mu}
\ee
where
\be
{\cal A}^{\mu}_{j}(p,q) = p^{\mu}~a_{j} + q^{\mu}_{j}~b_{j}, \quad j = 1,\dots,4
\ee
and
\bea
a_{1} = {q^{2} (p\cdot P)\over 2 N}, \quad b_{1} = - {p^{2} (q\cdot P)\over 2 N}
\nonumber\\
a_{2} = - {q\cdot P\over N}, \quad b_{2} = {p\cdot P\over N}
\nonumber\\
a_{3} = {p\cdot q\over N}, \quad b_{3} = - {p^{2} \over N}
\nonumber\\
a_{4} = {q^{2} \over N}, \quad b_{4} = - {p\cdot q\over N}.
\eea
In the limit 
$
N \rightarrow 0
$
the previous expression is continuous and we have
\be
\tilde{d}_{3}(p,q) = - i~(p - q)^{\mu}~F_{3} + i~P^{\mu}~(F_{1} + F_{2}).
\ee
\end{thm}
{\bf Proof:}
As in the previous Theorem, we obtain the first of the six contributions:
\be
\tilde{d}^{\mu(1)}_{3}(p,q) = - {i \over (2\pi)^{6}}~\int dk {k^{\mu} \over k^{2} - m^{2} - i~0}~
\theta(p_{0} - k_{0})~\delta((p - k)^{2} - m^{2})~\theta(k_{0} + q_{0})~\delta((k + q)^{2} - m^{2}).
\label{d1mu}
\ee
If we make the change of variables
$
k \rightarrow k + p
$
we obtain 
\be
\tilde{d}^{\mu(1)}(p,q) = - i~[ p^{\mu}~\tilde{d}^{(1)}(p,q) + e^{\mu}(p,q) ]
\label{d1mua}
\ee
where
\be
e^{\mu}(p,q) = {1 \over (2\pi)^{6}}~\int dk {k^{\mu} \over (k + p)^{2} - m^{2} - i~0}~
\theta(- k_{0})~\delta(k^{2} - m^{2})~\theta(k_{0} + P_{0})~\delta((k + P)^{2} - m^{2}).
\label{e}
\ee
We proceed as in the previous theorem and obtain as in (\ref{d1b})
\be
e^{\mu}(p,q) = {1 \over (2\pi)^{6}}~\int_{\omega_{\bf k} \leq P_{0}} {d{\bf k} \over 2 \omega_{\bf k}}~\tau^{\mu}({\bf k})~
\delta(P^{2} - 2 P_{0}\omega_{\bf k} - 2 {\bf P}\cdot {\bf k})~
(p^{2} - 2 p_{0}\omega_{\bf k} - 2 {\bf p}\cdot {\bf k} - i~0)^{-1}
\label{e1}
\ee
where
$
\tau^{\mu}({\bf k}) = (- \omega_{\bf k}, {\bf k}). 
$
Next, we use Lorentz covariance and do the computations in the particular frame we have used above; the result is (for 
$
P^{2} > 0,~P^{0} \geq 0 
$):
\be
e^{\mu}(p,q) = {1 \over (2\pi)^{6}}~\int_{\omega_{\bf k} \leq M} {d{\bf k} \over 2 M^{2}}~~\tau^{\mu}({\bf k})~
\delta\Bigl(\omega_{\bf k} - {M\over 2}\Bigl)~
(p^{2} - M p_{0} - 2 {\bf p}\cdot {\bf k} - i~0)^{-1}
\label{e2}
\ee
We consider the case
$
{\bf p} \not= {\bf 0}
$
and treat separately the cases
$
\mu = 0
$
and
$
\mu \not= 0.
$
The first case is easy:
\be
e^{0}(p,q) = - {1 \over 2}~M~\tilde{d}^{(1)}(p,q).
\ee
We also have
\be
e^{1} = e^{2} = 0.
\ee
The remaining case can be treated as in the preceding theorem; 
\be
e^{3}(p,q) = \theta(P_{0})~\theta(P^{2} - 4m^{2})~{r_{0}^{2} \over 4(2\pi)^{5}M}~I_{1}(A,B)
\label{e3}
\ee
where
\be
I_{1}(A,B) \equiv \int_{-1}^{1} {dz z\over A - B z}
\ee
and $A$ and $B$ have the same values as before:
$
A = p^{2} - M p_{0} - i~0,\quad B = 2 |{\bf p}|r_{0}.
$
The integral is elementary:
\be
I_{1}(A,B) = {1 \over B}~\Bigl[ - 2 + {A \over B}~ln\Bigl({A + B\over A - B}\Bigl)\Bigl].
\ee
In the case 
$
|{\bf p}| = 0
$
we easily obtain
\be
e^{3}(p,q) = 0.
\ee
Again, as in the previous theorem, we obtain that the limit 
$
|{\bf p}| \rightarrow 0
$
of (\ref{e3}) exists and is $0$. It remains to go to an arbitrary frame. After a tedious computation we obtain for 
$
N \not= 0
$
\be
\tilde{d}_{3}^{\mu(1)}(p,q) = i~({\cal A}^{\mu}_{1}~\tilde{d}^{(1)} + {\cal A}^{\mu}_{2}~\tilde{f}^{(+)}_{3})
\ee
where the expressions
$
{\cal A}_{j}~, j = 1,2
$
are those from the statement. For 
$
N = 0
$
we get
\be
\tilde{d}_{3}^{\mu(1)}(p,q) = - {i \over p\cdot q}~(p - q)^{\mu}~\tilde{f}^{(+)}_{3}
\ee
If we use now relations similar to (\ref{d2}) - (\ref{d6}) we get the other five contributions and the relation from the statement
follows.
$\qed$

The expression
$
\tilde{d}^{\mu}_{1}, \tilde{d}^{\mu}_{2}
$
can be obtained from
$
\tilde{d}^{\mu}_{3}
$
by clever changes of variables, as in \cite{loop}. We note that for 
$
N \not= 0
$
the expressions 
$
\tilde{d}^{\mu}_{j}
$
obtained above are identical to those from \cite{loop} where the derivation was made by another method.
\newpage
Finally we define
\be
d_{jk}^{\mu\nu} \equiv {\cal D}_{j}^{\mu} {\cal D}_{k}^{\nu}d
\label{dij}
\ee
and we have the following orders of singularity:
\be
\quad \omega(d_{jk}^{\mu\nu}) = 0.
\label{deg2}
\ee
We will first consider the case 
$
d_{33}.
$
The result is
\begin{thm}
For 
$
N \not= 0
$
the following formula is true:
\be
\tilde{d}_{33}^{\mu\nu}(p,q) = {\cal A}^{\mu\nu}_{1}~\tilde{d} + {\cal A}^{\mu\nu}_{2}~\tilde{f}_{3} + 
{\cal A}^{\mu\nu}_{3}~\tilde{f}_{1} + {\cal A}^{\mu\nu}_{4}~\tilde{f}_{2}
\label{d3munu}
\ee
where
\be
{\cal A}^{\mu\nu}_{j}(p,q) = - [ p^{\mu} p^{\nu}~\alpha_{j} + q^{\mu} q^{\nu}~\beta_{j}
+ (p^{\mu} q^{\nu} + p^{\nu} q^{\mu})~\gamma_{j} + \eta^{\mu\nu}~\delta_{j}], \quad j = 1,\dots,4
\ee
and
\bea
\alpha_{1} = {3 P^{2} p^{2} (q^{2})^{2} \over 8 N^{2}} + {(q^{2})^{2} \over 4 N} + {m^{2} q^{2} \over 2 N},
\quad \beta_{1} = {3 P^{2} q^{2} (p^{2})^{2} \over 8 N^{2}} + {(p^{2})^{2} \over 4 N} + {m^{2} p^{2} \over 2 N}
\nonumber\\
\gamma_{1} = - {3 P^{2} p^{2} q^{2} (p\cdot q) \over 8 N^{2}} + {p^{2} q^{2} \over 4 N} - {m^{2} (p\cdot q) \over 2 N},
\quad \delta_{1} = {P^{2} p^{2} q^{2} \over 8 N} + {m^{2} \over 2}
\nonumber\\
\alpha_{2} = - {3 (P\cdot q)^{2} (p\cdot q) \over 4 N^{2}} + {4 P\cdot q + p\cdot q \over 4 N}, \quad
\beta_{2} = - {3 (P\cdot p)^{2} (p\cdot q) \over 4 N^{2}} + {4 P\cdot p + p\cdot q \over 4 N} 
\nonumber\\
\gamma_{2} = {3 (P\cdot p) (P\cdot q)  (p\cdot q) \over 4 N^{2}} - {2 P^{2} - p\cdot q \over 4 N}, \quad
\delta_{2} = - {P^{2} (p\cdot q) \over 4 N}
\nonumber\\
\alpha_{3} = {3 (p\cdot q)^{2} (P\cdot q) \over 4 N^{2}} - {4 p\cdot q + P\cdot q \over 4 N}, \quad
\beta_{3} = {3 (P\cdot q) (p^{2})^{2} \over 4 N^{2}} 
\nonumber\\
\gamma_{3} = - {3 (P\cdot q) (p\cdot q)  p^{2} \over 4 N^{2}} + {p^{2} \over 2 N}, \quad
\delta_{3} = {p^{2} (P\cdot q) \over 4 N}
\eea
and the expressions
$
\alpha_{4},\dots,\delta_{4}
$
are obtained from
$
\alpha_{3},\dots,\delta_{3}
$
making 
$
p \leftrightarrow q.
$
In the limit 
$
N \rightarrow 0
$
the previous expression is continuous and we have
\bea
\tilde{d}_{33}^{\mu\nu}(p,q) = - [ \alpha_{33}(p,q) P^{\mu} P^{\nu} + \eta^{\mu\nu}~\beta_{33}(p,q) ]F_{3} 
\nonumber\\
- [ \alpha_{33}(q,- P) p^{\mu} p^{\nu} + \eta^{\mu\nu}~\beta_{33}(q, - P) ]F_{1}
- [ \alpha_{33}(- p,P) q^{\mu} q^{\nu} + \eta^{\mu\nu}~\beta_{33}(- p,P) ]F_{2} 
\eea
where
\bea
\alpha_{33}(p,q) = {1 \over 6}~\Bigl[ 4 - {m^{2} \over 4 P^{2}} - 12 { (P\cdot p) (P\cdot q) \over (P^{2})^{2}} \Bigl]
\nonumber\\
\beta_{33}(p,q) = {P^{2} \over 6}~\Bigl( 1 - {m^{2} \over 4 P^{2}} \Bigl)
\eea
\end{thm}
{\bf Proof:}
As in the previous Theorems, we obtain the first of the six contributions:
\be
\tilde{d}^{\mu\nu(1)}(p,q) = - {1 \over (2\pi)^{6}}~\int dk {k^{\mu} k^{\nu} \over k^{2} - m^{2} - i~0}~
\theta(p_{0} - k_{0})~\delta((p - k)^{2} - m^{2})~\theta(k_{0} + q_{0})~\delta((k + q)^{2} - m^{2}).
\label{d1mnuu}
\ee
If we make the change of variables
$
k \rightarrow k + p
$
we obtain 
\be
\tilde{d}^{\mu\nu(1)}(p,q) = - p^{\mu} p^{\nu}~\tilde{d}^{(1)}(p,q) - [ p^{\mu} e^{\nu}(p,q) +  p^{\nu} e^{\mu}(p,q)]
- e^{\mu\nu}(p,q)
\label{d1munua}
\ee
where the expressions
$
e^{\mu}(p,q)
$
have been defined before - rel (\ref{e}) and
\be
e^{\mu\nu}(p,q) = {1 \over (2\pi)^{6}}~\int dk {k^{\mu}k^{\nu} \over (k + p)^{2} - m^{2} - i~0}~
\theta(- k_{0})~\delta(k^{2} - m^{2})~\theta(k_{0} + P_{0})~\delta((k + P)^{2} - m^{2}).
\label{ee}
\ee
We proceed as in the previous theorem and obtain as in (\ref{d1b})
\be
e^{\mu\nu}(p,q) = {1 \over (2\pi)^{6}}~\int_{\omega_{\bf k} \leq P_{0}} {d{\bf k} \over 2 \omega_{\bf k}}~
\tau^{\mu}({\bf k})~\tau^{\nu}({\bf k})~
\delta(P^{2} - 2 P_{0}\omega_{\bf k} - 2 {\bf P}\cdot {\bf k})~
(p^{2} - 2 p_{0}\omega_{\bf k} - 2 {\bf p}\cdot {\bf k} - i~0)^{-1}
\label{ee1}
\ee
where
$
\tau^{\mu}({\bf k}) = (- \omega_{\bf k}, {\bf k}). 
$
Next, we use Lorentz covariance and do the computations in the particular frame we have used above; the result is:
\be
e^{\mu\nu}(p,q) = {1 \over (2\pi)^{6}}~\int_{\omega_{\bf k} \leq M} {d{\bf k} \over 2 M^{2}}~~\tau^{\mu}({\bf k})~\tau^{\nu}({\bf k})~
\delta\Bigl(\omega_{\bf k} - {M\over 2}\Bigl)~
(p^{2} - M p_{0} - 2 {\bf p}\cdot {\bf k} - i~0)^{-1}
\label{ee2}
\ee
We consider the case
$
{\bf p} \not= {\bf 0}.
$
We easily obtain
\be
e^{00}(p,q) = {1 \over 4}~M^{2}~\tilde{d}^{(1)}(p,q), \quad e^{\mu 0}(p,q) = - {1 \over 2}~M~e^{\mu}(p,q)
\ee
We also have
\be
e^{jk} = 0,~j, k = 1,2,3,~j \not= k.
\ee
Next 
\be
e^{33}(p,q) = \theta(P_{0})~\theta(P^{2} - 4m^{2})~{r_{0}^{3} \over 4(2\pi)^{5}M}~I_{2}(A,B)
\ee
where
\be
I_{2}(A,B) \equiv \int_{-1}^{1} {dz z^{2}\over A - B z}
\ee
and $A$ and $B$ have the same values as before:
$
A = p^{2} - M p_{0} - i~0,\quad B = 2 |{\bf p}|r_{0}.
$
The integral is elementary:
\be
I_{2}(A,B) = {A \over B}~I_{1}(A,B)
\ee
In the case 
$
|{\bf p}| = 0
$
the expression
$
e^{33}(p,q)
$
is the limit 
$
|{\bf p}| \rightarrow 0
$
of the previous expression. The expressions
$
e^{11}
$
and
$
e^{22}
$
can be obtained similarly. It remains to to an arbitrary frame. After a tedious computation we obtain for 
$
N \not= 0
$
\be
\tilde{d}_{3}^{\mu\nu(1)}(p,q) = {\cal A}^{\mu\nu}_{1}~\tilde{d}^{(1)} + {\cal A}^{\mu\nu}_{2}~\tilde{f}^{(+)}_{3}
\ee
where the expressions
$
{\cal A}_{j}~, j = 1,2
$
are those from the statement. For 
$
N = 0
$
we get
\be
\tilde{d}_{3}^{\mu\nu(1)}(p,q) = - (P^{\mu} P^{\nu} \alpha + \eta^{\mu\nu} \beta)
\ee
where
\bea
\alpha = - {1 \over 6 p\cdot q}~\Bigl[ 4 - {m^{2} \over 4 P^{2}} - 12 { (P\cdot p) (P\cdot q) \over (P^{2})^{2}} \Bigl]f^{(+)}_{3} 
\nonumber\\
\beta = - {P^{2} \over 6 p\cdot q}~\Bigl( 1 - {m^{2} \over 4 P^{2}} \Bigl)f^{(+)}_{3}
\eea
If we use now relations similar to (\ref{d2}) - (\ref{d6}) we get the other five contributions and the relation from the statement
follows.
$\qed$

The expression
$
\tilde{d}^{\mu}_{11}, \tilde{d}^{\mu}_{22}
$
can be obtained from
$
\tilde{d}^{\mu}_{33}
$
by clever changes of variables, as in \cite{loop}. We note that for 
$
N \not= 0
$
the expressions 
$
\tilde{d}^{\mu}_{jj}
$
obtained above are identical to those from \cite{loop} where the derivation was made by another method.

We still have to consider the case 
$
d_{12}^{\mu\nu}.
$
The result is
\begin{thm}
For 
$
N \not= 0
$
the following formula is true:
\be
\tilde{d}_{12}^{\mu\nu}(p,q) = {\cal B}^{\mu\nu}_{1}~\tilde{d} + {\cal B}^{\mu\nu}_{2}~\tilde{f}_{3} + 
{\cal B}^{\mu\nu}_{3}~\tilde{f}_{1} + {\cal B}^{\mu\nu}_{4}~\tilde{f}_{2}
\ee
where
\be
{\cal B}^{\mu\nu}_{j}(p,q) = p^{\mu} p^{\nu}~A_{j} + q^{\mu} q^{\nu}~B_{j}
+ p^{\mu} q^{\nu} C_{j}^{(1)} + p^{\nu} q^{\mu}~C_{j}^{(2)} + \eta^{\mu\nu}~D_{j}, \quad j = 1,\dots,4
\ee
and
\bea
A_{1} = {3 P^{2} p^{2} (q^{2})^{2} \over 8 N^{2}} + {(q^{2})^{2} \over 4 N} + {q^{2} (P\cdot p) \over 4 N} + {m^{2} q^{2} \over 2 N},
\nonumber\\
B_{1} = {3 P^{2} q^{2} (p^{2})^{2} \over 8 N^{2}} + {(p^{2})^{2} \over 4 N} + {p^{2} (P\cdot q) \over 4 N} + {m^{2} p^{2} \over 2 N},
\nonumber\\
C_{1}^{(1)} = - {3 P^{2} p^{2} q^{2} (p\cdot q) \over 8 N^{2}} + {p^{2} q^{2} \over 4 N} - {m^{2} p\cdot q \over 2 N},
\nonumber\\
C_{1}^{(2)} = - {3 P^{2} p^{2} q^{2} (p\cdot q) \over 8 N^{2}} + {p^{2} q^{2} \over 4 N} 
- {P^{2} (p\cdot q) \over 2 N} - {m^{2} (p\cdot q) \over 2 N},\quad 
D_{1} = {P^{2} p^{2} q^{2} \over 8 N} + {m^{2} \over 2}
\nonumber\\
A_{2} = - {3 (P\cdot q)^{2} (p\cdot q) \over 4 N^{2}} + {p\cdot q \over N}, \quad 
B_{2} = - {3 (P\cdot p)^{2} (p\cdot q) \over 4 N^{2}} + {p\cdot q \over N}
\nonumber\\
C_{2}^{(1)} = {3 (P\cdot p) (P\cdot q) (p\cdot q) \over 4 N^{2}} - {P^{2} \over 2 N} + {p\cdot q \over 4 N},\quad 
\nonumber\\
C_{2}^{(2)} = {3 (P\cdot p) (P\cdot q) (p\cdot q) \over 4 N^{2}} + {P^{2} \over 2 N} + {p\cdot q \over 4 N}, \quad
D_{2} = - {P^{2} (p\cdot q) \over 4 N}
\nonumber\\
A_{3} = {3 (p\cdot q)^{2} (P\cdot q) \over 4 N^{2}} - {P\cdot q \over 4 N}, \quad 
B_{3} = {3 (p^{2})^{2} (P\cdot q) \over 4 N^{2}} + {p^{2} \over N}
\nonumber\\
C_{3}^{(1)} = - {3 (p\cdot q) (P\cdot q) p^{2} \over 4 N^{2}} + {p^{2} \over 2 N},\quad 
\nonumber\\
C_{3}^{(2)} = - {3 (p\cdot q) (P\cdot q) p^{2} \over 4 N^{2}} - {p^{2} \over 2 N} - {p\cdot q \over N},\quad
D_{3} = {p^{2} (P\cdot q) \over 4 N}
\nonumber\\
A_{4} = {3 (q^{2})^{2} (P\cdot p) \over 4 N^{2}} + {q^{2} \over N}, \quad 
B_{4} = {3 (p\cdot q)^{2} (P\cdot p) \over 4 N^{2}} - {p\cdot P \over 4 N}
\nonumber\\
C_{4}^{(1)} = - {3 (p\cdot q) (P\cdot p) q^{2} \over 4 N^{2}} + {q^{2} \over 2 N},\quad 
\nonumber\\
C_{4}^{(2)} = - {3 (p\cdot q) (P\cdot p) q^{2} \over 4 N^{2}} - {q^{2} \over 2 N} - {p\cdot q \over N},\quad 
D_{4} = {q^{2} (P\cdot p) \over 4 N}
\eea
In the limit 
$
N \rightarrow 0
$
the previous expression is continuous and we have
\bea
\tilde{d}_{12}^{\mu\nu}(p,q) = [ \alpha_{12}(p,q) P^{\mu} P^{\nu} + \eta^{\mu\nu}~\beta_{12}(p,q) ]F_{3} 
\nonumber\\
+ [ \alpha_{12}(q,- P) p^{\mu} p^{\nu} + \eta^{\mu\nu}~\beta_{12}(q, - P) ]F_{1}
+ [ \alpha_{12}(- p,P) q^{\mu} q^{\nu} + \eta^{\mu\nu}~\beta_{12}(- p,P) ]F_{2} 
\eea
where
\bea
\alpha_{12}(p,q) = - {1 \over 6}~\Bigl( 2 + {m^{2} \over 4 P^{2}} \Bigl), \quad 
\beta_{12}(p,q) = {P^{2} \over 6}~\Bigl( 1 - {m^{2} \over 4 P^{2}} \Bigl).
\eea
\end{thm}
{\bf Proof:}
As in the previous Theorems, we obtain the first of the six contributions:
\be
\tilde{d}^{\mu\nu(1)}_{12}(p,q) = - {1 \over (2\pi)^{6}}~\int dk {(k + q)^{\mu} (k - p)^{\nu} \over k^{2} - m^{2} - i~0}~
\theta(p_{0} - k_{0})~\delta((p - k)^{2} - m^{2})~\theta(k_{0} + q_{0})~\delta((k + q)^{2} - m^{2}).
\label{d12mnuu}
\ee
If we make the change of variables
$
k \rightarrow k + p
$
we obtain 
\be
\tilde{d}^{\mu\nu(1)}_{12}(p,q) = P^{\mu} e^{\nu}(p,q) + e^{\mu\nu}(p,q)
\label{d12munua}
\ee
where the expressions
$
e^{\mu}(p,q)
$
and 
$
e^{\mu\nu}
$
have been defined before - rel (\ref{e}) and (\ref{ee}). Proceeding as before we get the formulas from the statement.
$\qed$

The expression
$
\tilde{d}^{\mu}_{23}, \tilde{d}^{\mu}_{31}
$
can be obtained from
$
\tilde{d}^{\mu}_{12}
$
by clever changes of variables, as in \cite{loop}. We note that for 
$
N \not= 0
$
the expressions 
$
\tilde{d}^{\mu}_{jk},~j \not= k
$
obtained above are identical to those from \cite{loop} where the derivation was made by another method.

One can obtain in the same way the expressions
\be
d_{jkl}^{\mu\nu\rho} \equiv {\cal D}_{j}^{\mu} {\cal D}_{k}^{\nu} {\cal D}_{l}^{\rho}d.
\label{dijk}
\ee

We only emphasize in the end the main idea: the chronological products can be obtained from the preceding theorems by a simple 
operation, namely the causal splitting of a master distribution $d$ given by (\ref{d123}). An explicit procedure to do this
causal splitting can be found in \cite{Sc1} and \cite{Sc2}. In fact, if we want to split causally (\ref{d3mu}) it is better to
multipy it by $N$ so, if we go in the coordinate spsce, we will have in both hand sides some polynomials in the partial derivatives 
acting on distributions with causal support. The same idea is valid for (\ref{d3munu}), but we have to multipy by 
$
N^{2}.
$

\end{document}